\documentclass[12pt]{article}
\usepackage[a4paper, margin=2.5cm]{geometry} 
\usepackage[hidelinks]{hyperref}
\usepackage{xcolor}

\hypersetup{
    colorlinks=true,    
    linkcolor=black,    
    citecolor=black,    
    urlcolor=black,     
}

\usepackage{mathpazo}
\usepackage{subfig}
\usepackage{subcaption}
\usepackage{graphicx}         
\usepackage{booktabs}         
\usepackage{amsmath,amssymb}  
\usepackage{hyperref}         
\usepackage[separate-uncertainty=true]{siunitx}

\title{Measurements of electroweak penguin and lepton-flavor violating $B$~decays to final states with missing energy at Belle and Belle~II}
\author{Gaetano de Marino\thanks{Jo\v{z}ef Stefan Institute, Jamova cesta 39, 1000 Ljubljana. Email: \href{mailto:gaetano.demarino@ijs.s}{gaetano.demarino@ijs.si}}, on behalf of the Belle~II collaboration}

\date{} 

\begin{document}

\maketitle
\begin{center}
\textbf{{\large Presented at the 32$\boldsymbol{^{\rm nd}}$ International Symposium on Lepton Photon Interactions at High Energies, Madison, Wisconsin, USA, August~25-29,~2025}}
\end{center}
\vspace{0.8cm}
\begin{abstract}
The Belle and Belle II experiments have collected a $\SI{1.2}{ab}^{-1}$ sample of collisions at a center-of-mass energy corresponding to the $\Upsilon(4S)$ resonance. These datasets, with low particle multiplicity and constrained initial state kinematics, are an ideal environment to search for rare electroweak penguin $B$~decays and lepton-flavor-violating $B$~decays to final states with missing energy from neutrinos. Results from $b \to s\nu\bar{\nu}$ processes and their interpretation are presented. In addition, we provide an overview of the search for the $B \to K^{*0}\tau^+\tau^-$ decays and the lepton-flavor violating decays $B^0 \to K^{(*)0}\tau^\pm\ell^\mp$, where $\ell$ is an electron or a muon.
\end{abstract}
\vspace{0.6cm}
\section{Introduction}
$B$ meson decays are exhibiting, at different levels of significance, discrepancies -- or \textit{anomalies} -- with respect to the Standard Model (SM) predictions, which might hint at common New Physics (NP) effects~\cite{knunu,c9fitLHCb,HFLAV:Spring25}. In flavor-changing neutral current processes, beyond-SM particles can contribute through additional diagrams where they appear as mediators, either at tree or loop level, or in the final state. Given the current experimental status, final states involving third-generation leptons ($\tau$, $\nu_\tau$) are expected to couple more strongly to NP particles potentially related to these anomalies. However, the presence of undetectable particles (missing energy) makes the corresponding searches and measurements particularly challenging. \\These proceedings summarize recent published and preliminary results obtained using the data collected by the Belle and Belle~II experiments. The transitions $b \to s\nu\bar{\nu}$, $b \to s\tau\bar{\tau}$, and $b \to s\tau\bar{\ell}$ are discussed in dedicated sections below.

\section{B-factories and missing energy}
The $B$-factories KEKB and its upgrade SuperKEKB have collected $e^+e^-$ collision data at the $\Upsilon(4S)$ resonance corresponding to $\SI{711}{fb}^{-1}$ (Belle) and $365\,\text{fb}^{-1}$ (Run~1)~+~$125\,\text{fb}^{-1}$ (Run~2) for Belle~II. The results presented here use either Belle~II-only or combined Belle and Belle~II samples, excluding Run~2 data.

The $\Upsilon(4S)$ decays almost exclusively into $B^0\bar{B}^0$ or $B^+B^-$ pairs. For analyses involving missing energy, a reconstruction of the companion $B$ meson ($B$-tagging) is performed to interpret the event and infer the otherwise inaccessible signal-$B$ properties.
In the inclusive tagging approach, all reconstructed particles other than the signal are used to define the tag-$B$, a method well suited for low-multiplicity decays such as $B\to K\nu\bar{\nu}$. Other analyses employ hadronic tagging, in which the tag-$B$ candidates are fully reconstructed in known hadronic modes. This method provides precise event kinematics knowledge but suffers from low efficiency, which is mitigated using multivariate algorithms~\cite{fei}.

\section{\texorpdfstring{$\boldsymbol{b\to s \nu\bar{\nu}}$}{btoknunu} searches}
\begin{figure}[!ht]
\centering
\subfloat[]{\includegraphics[width=0.48\linewidth]{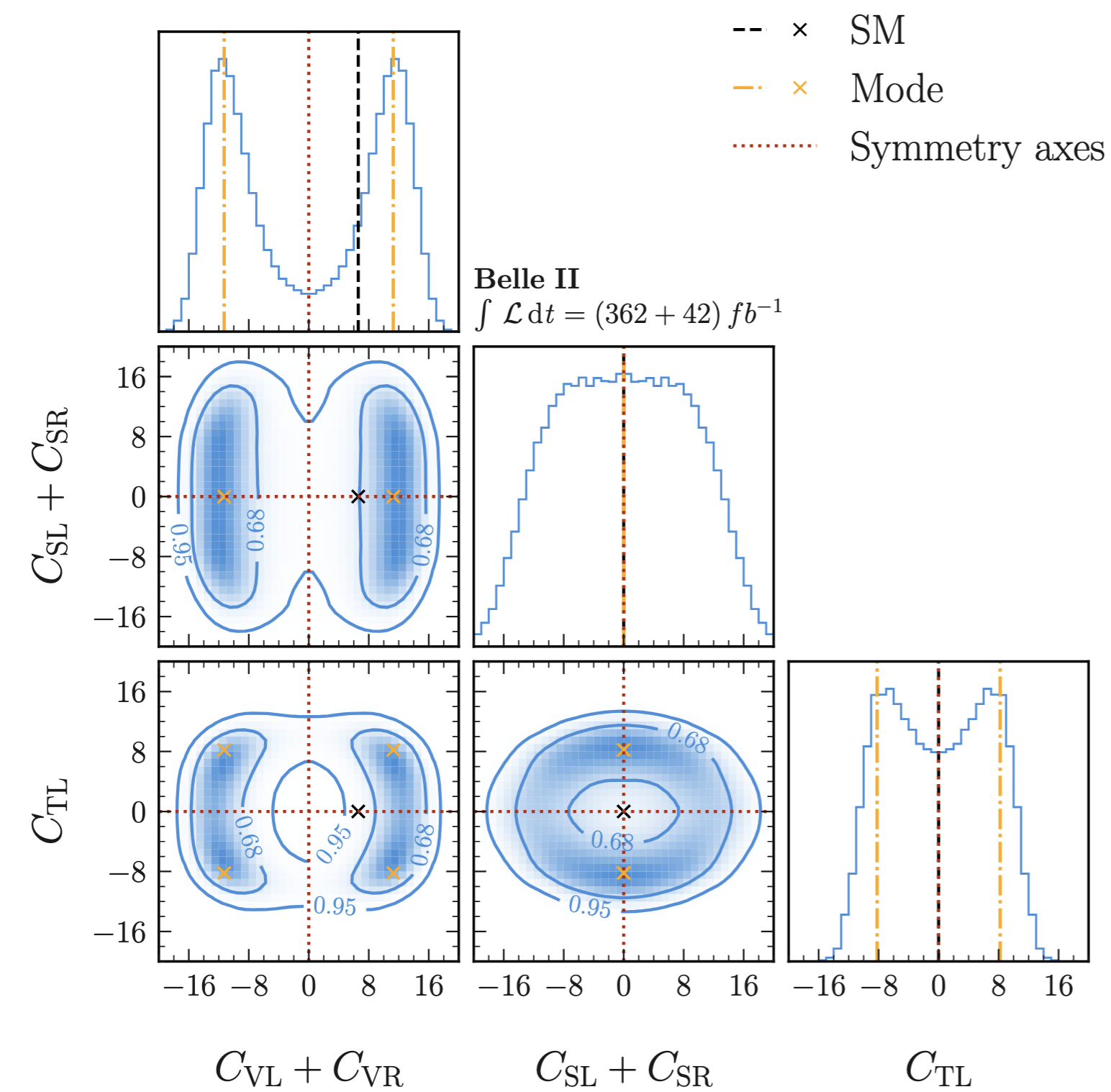}\label{fig:a}}
\quad
\subfloat[]{\includegraphics[width=0.48\linewidth]{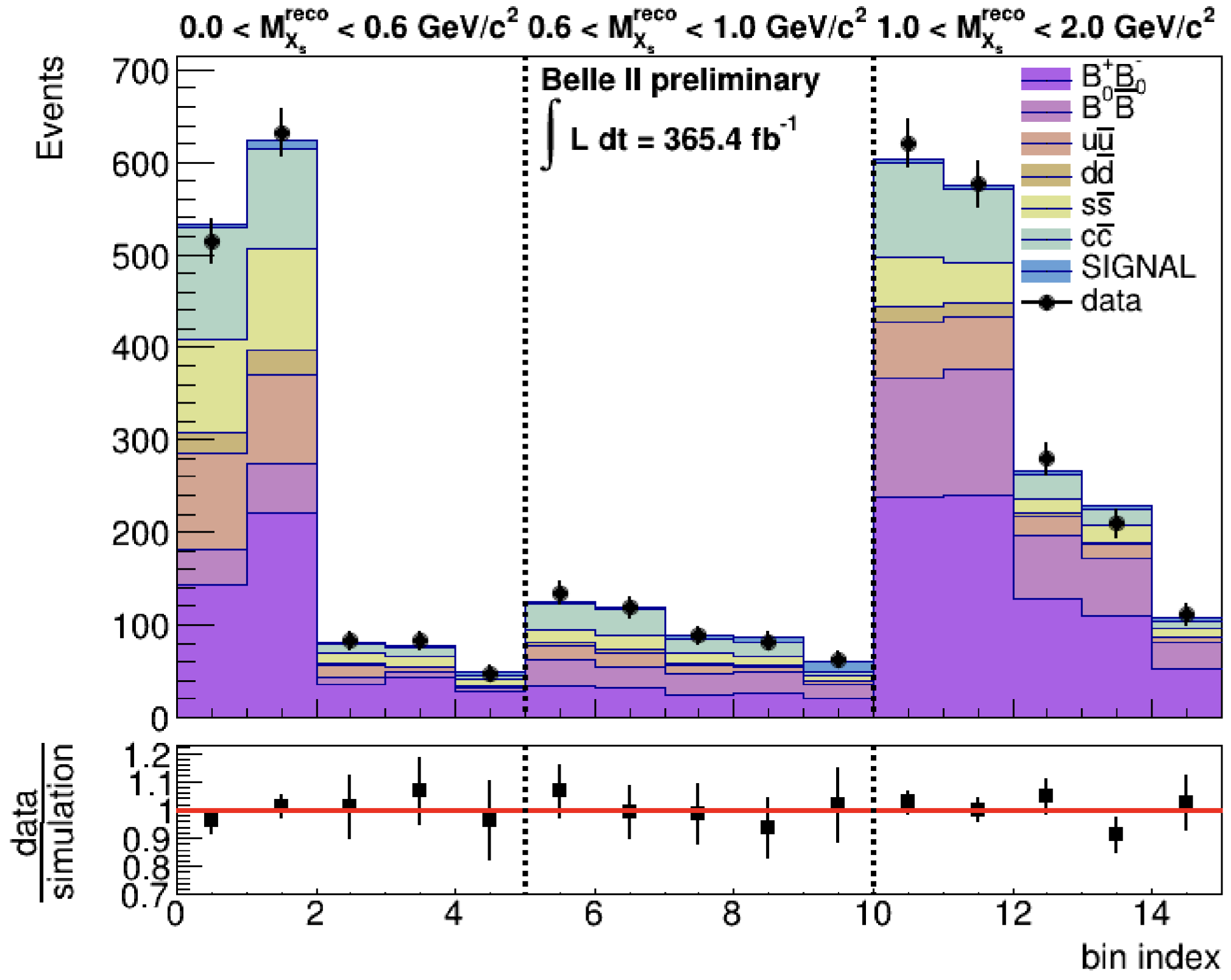}\label{fig:b}}
\caption{$(b \to s\nu\bar{\nu})$ recent Belle~II results. \protect\subref{fig:a}: Reinterpretation~\cite{reint} of the $B^+ \to K^+\nu\bar{\nu}$ evidence with marginalized posterior for WET Wilson coefficients (real $C_{V_R,S_R,T_R}$ sums). Contours: 68/95\% credible intervals; dashed: SM; dash-dotted: mode; dotted: symmetry axes for sample symmetrization. \protect\subref{fig:b}: $B\to X_s\nu\bar{\nu}$ data and post-fit histogram templates in three $X_s$ mass regions, with bins representing optimized BDT output ranges. Backgrounds are separated into charged and neutral B decays, and four $e^+e^- \to q\bar{q}$ components ($q = u, d, s, c)$. \label{fig:snunu}}
\end{figure}

The first evidence of the $B^+ \to K^+\nu\bar{\nu}$ decays reported by Belle~II~\cite{knunu} assumes the Standard Model (SM) signal expectation~\cite{Parrott}. To further investigate the observed $2.7~\sigma$ excess, a reinterpretation using a histogram reweighting method~\cite{lorenz} is performed for the first time. With the null hypothesis being built from the SM cross section, alternative models can be probed by applying the appropriate weights. For $B^+ \to K^+\nu\bar{\nu}$ decays, these weights depend on $q^2=(p_\nu+p_{\bar{\nu}})^2$, mapped to the reconstructed variable\footnote{$s$ is the squared center-of-mass (c.m.) energy, $M_K$ the known $K^+$ mass, and $E_K^*$ the reconstructed kaon energy in the c.m. frame.} $q^2_{\rm rec} = s/(4c^4)+M_K^2-\sqrt{s}E_K^*/c^4$, together with the signal classifier output. This approach enables quantitative inference on alternative scenarios, including those with additional dimension-six operators in the weak effective theory (WET) framework. The data favor enhanced vector and nonzero tensor contributions compared to the SM prediction (Fig.~\ref{fig:a}). The paper~\cite{reint} shows the feasibility of the method and provides the likelihood inputs for ready-to-use reinterpretations.

We have recently performed a search for $B \to X_s\nu\bar{\nu}$ decays using a sum-of-exclusive approach and Belle~II data. This technique, applied for the first time to decays with missing energy, improves upon the only previous measurement~\cite{ALEPH}, whose upper limit remains well above the precise SM prediction~\cite{Buras}. The analysis employs hadronic $B$-tagging, followed by reconstruction of the $X_s$ system in 30 modes with strangeness $S=\pm1$.
Background suppression relies on a multivariate algorithm (boosted decision trees, BDT's), where strong discrimination is provided by the residual energy in the electromagnetic calorimeter originating from unreconstructed particles or background.
The fit to the data (Fig.~\ref{fig:b}) is performed in three different $X_s$ reconstructed invariant mass ranges, which are enhanced in the $K$, $K^*(892)$ and higher-mass $X_s$ modes. For signal extraction, the reconstructed mass is converted back to the true one $M(X_s)$. 
The branching fractions measured in the three $X_s$ mass regions are not statistically significant, and upper limits are set. The result in the $K$ mass region agrees with that of Ref.~\cite{knunu} (hadronic tag). The combined 90\% confidence level limit on $\mathcal{B}(B \to X_s\nu\bar{\nu})$ is the most stringent obtained so far.

\section{\texorpdfstring{$\boldsymbol{b\to s \tau\bar{\tau}}$}{btoktautau} searches}

\begin{figure}[!ht]
    \centering
        \includegraphics[width=\linewidth]{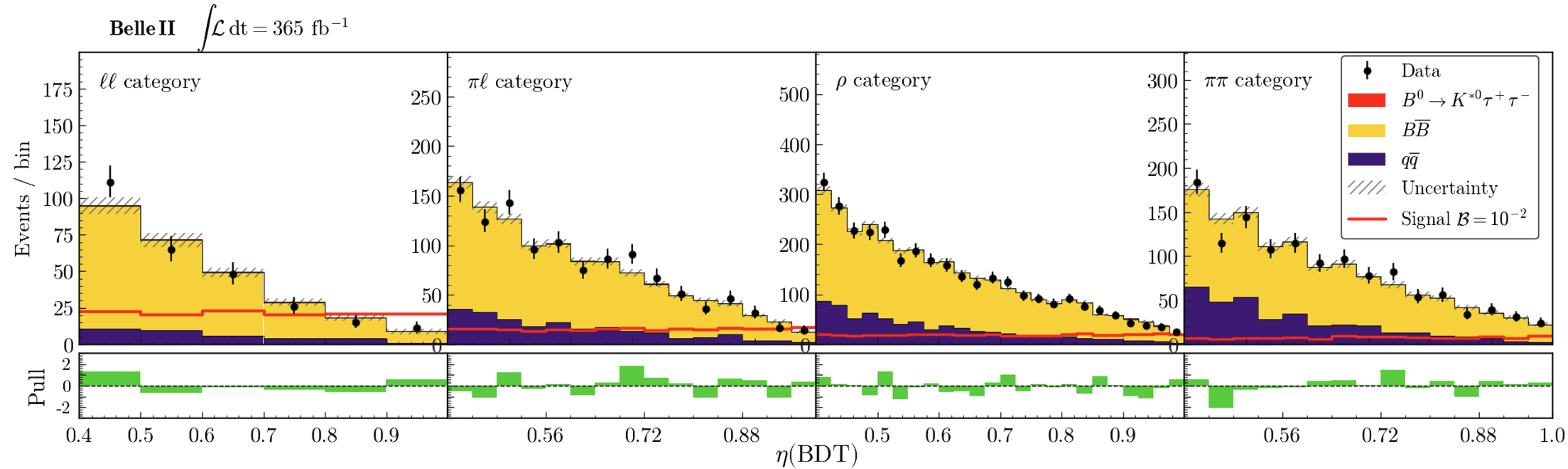}
        \caption{Transformed BDT distributions for four signal categories, showing $B\bar{B}$ and $q\bar{q}\;(q=u,d,s,c)$ backgrounds, the $B^0 \to K^{0}\tau^+\tau^-$ signal, and a reference signal at $\mathcal{B}=10^{-2}$. From~\cite{K*0tautau}.\label{fig:stautau}}
\end{figure}

The $b\to s\tau^+\tau^-$ decays have SM branching fractions of $\mathcal{O}(10^{-7})$, far below current experimental sensitivity. However, they can be significantly enhanced — by up to three orders of magnitude in new physics scenarios addressing the anomalies in $R(D^{(*)})$ and $\mathcal{B}(B^+\to K^+\nu\bar{\nu})$. At Belle~II, we search for $B^0 \to K^{*0}\tau^+\tau^-$ decays~\cite{K*0tautau} using the hadronic $B$-tagging approach, reconstructing both $\tau$ leptons in their dominant one-prong decay modes. A BDT is employed to suppress backgrounds, exploiting missing-energy information, reconstructed $q^2$, kinematic variables, and $K^{*0}$ properties. Calibration and validation are performed using off-resonance data\footnote{Collected about $\SI{60}{MeV}$ below the $\Upsilon(4S)$ resonance, providing a $B\bar{B}$-free sample.}, same-flavor events, and clean $B^0\to K^{*0}J/\psi$ decays. The signal yield is extracted from a fit to the BDT output in a signal-enhanced region (Fig.~\ref{fig:stautau}), performed simultaneously across four di-$\tau$ final-state categories with different purities and background compositions. No significant signal is observed, and an upper limit on the branching fraction is set, improving substantially over the previous Belle result~\cite{bdtokstartautau}.

\section{\texorpdfstring{$\boldsymbol{b\to s \tau\{e\,\mu\}}$}{btoktaul} searches}
Charged lepton flavor violation (cLFV) is allowed in the SM via neutrino mixing, but at rates far below experimental sensitivity. Several NP models addressing tensions in $B$~decays, while respecting other flavor constraints, can produce measurable cLFV rates~\cite{Allwicher}. The experimental sensitivity for $b \to s\tau\ell$ decays has now reached the sub-$10^{-5}$ level with Belle~\cite{Kptauell} and LHCb~\cite{Ksztaue}.

\begin{figure}
\centering
\subfloat[]{\includegraphics[width=0.31\linewidth]{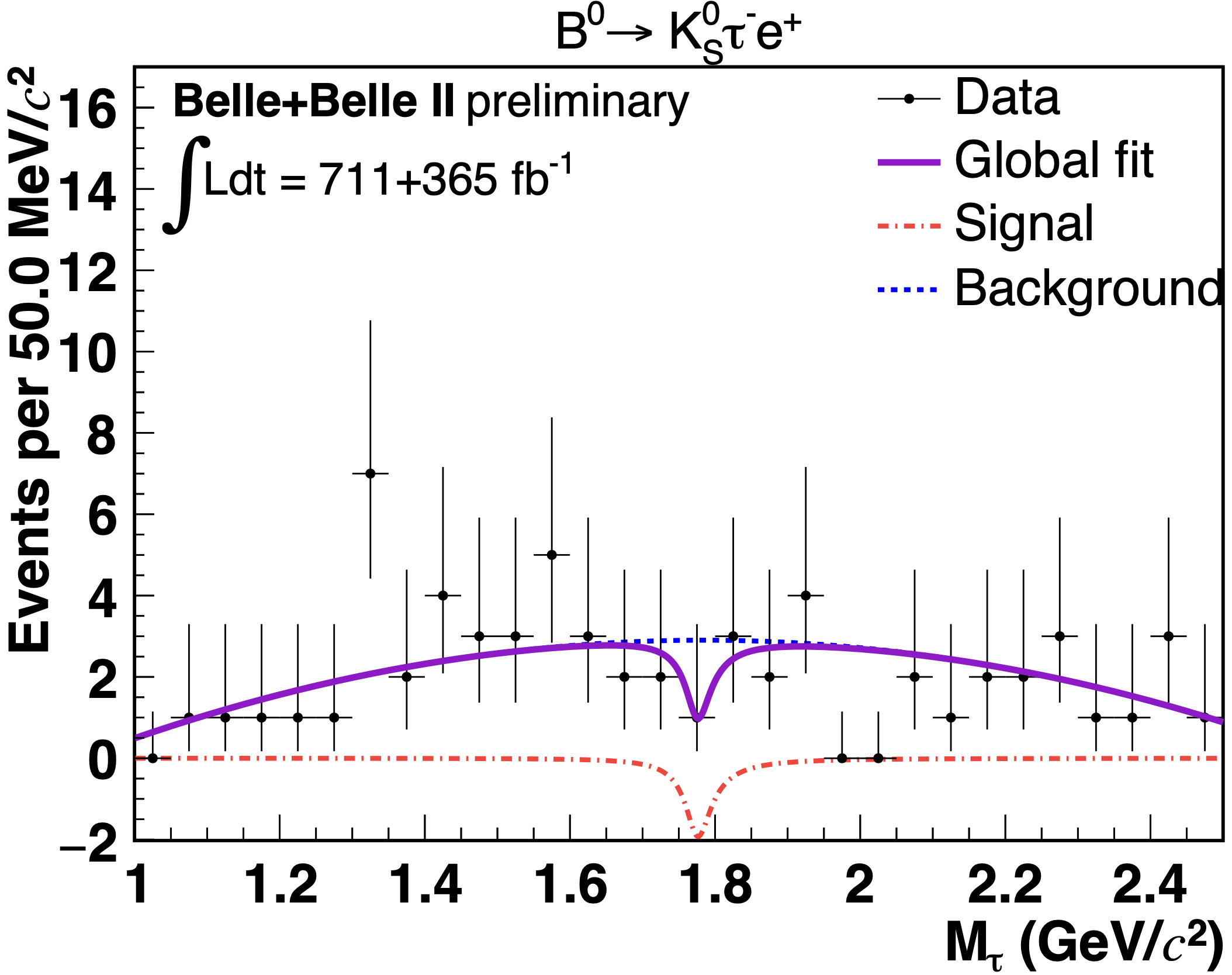}\label{fig:a2}}
\quad
\subfloat[]{\includegraphics[width=0.31\linewidth]{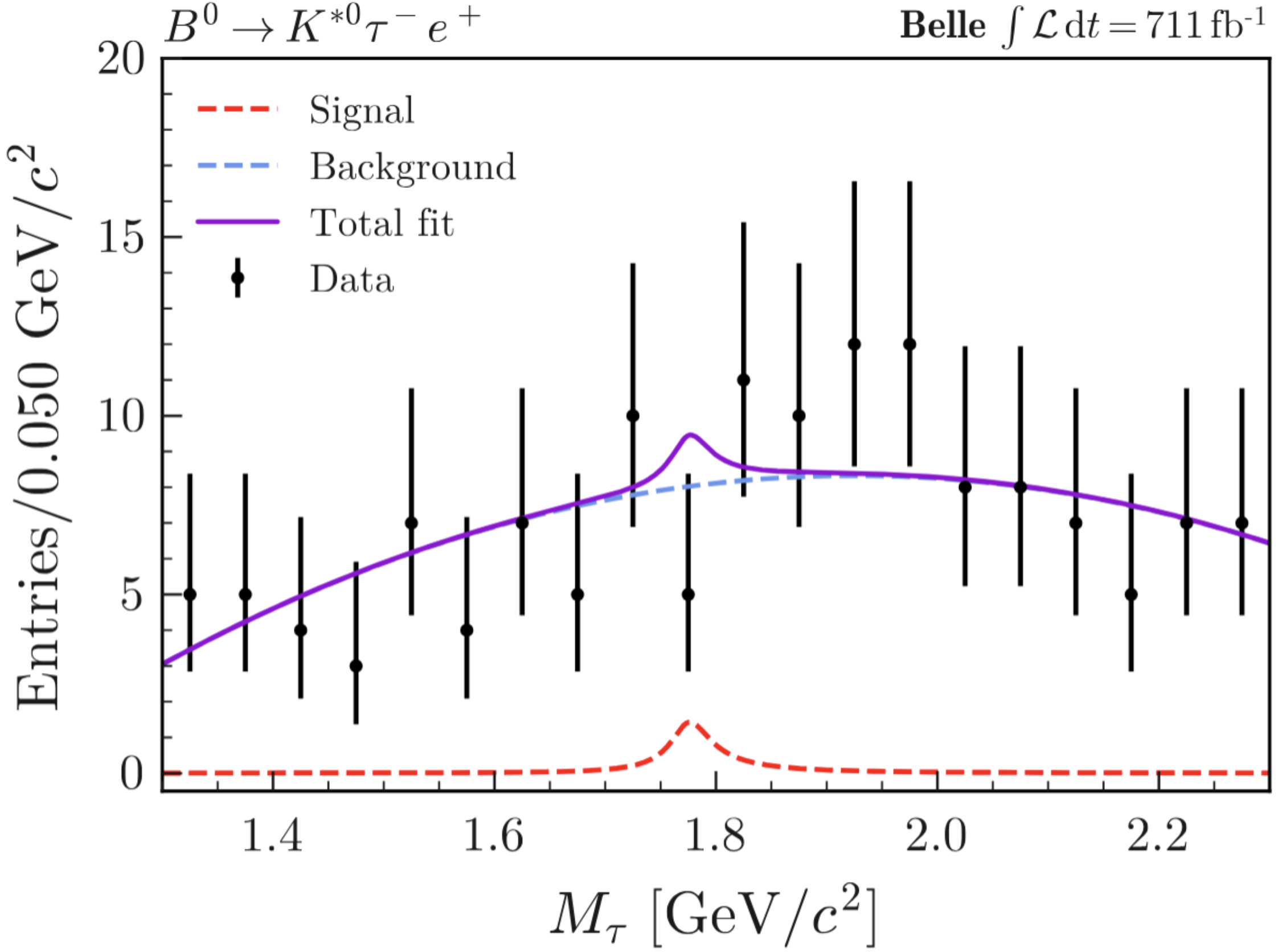}\label{fig:b2}}
\quad
\subfloat[]{\includegraphics[width=0.31\linewidth]{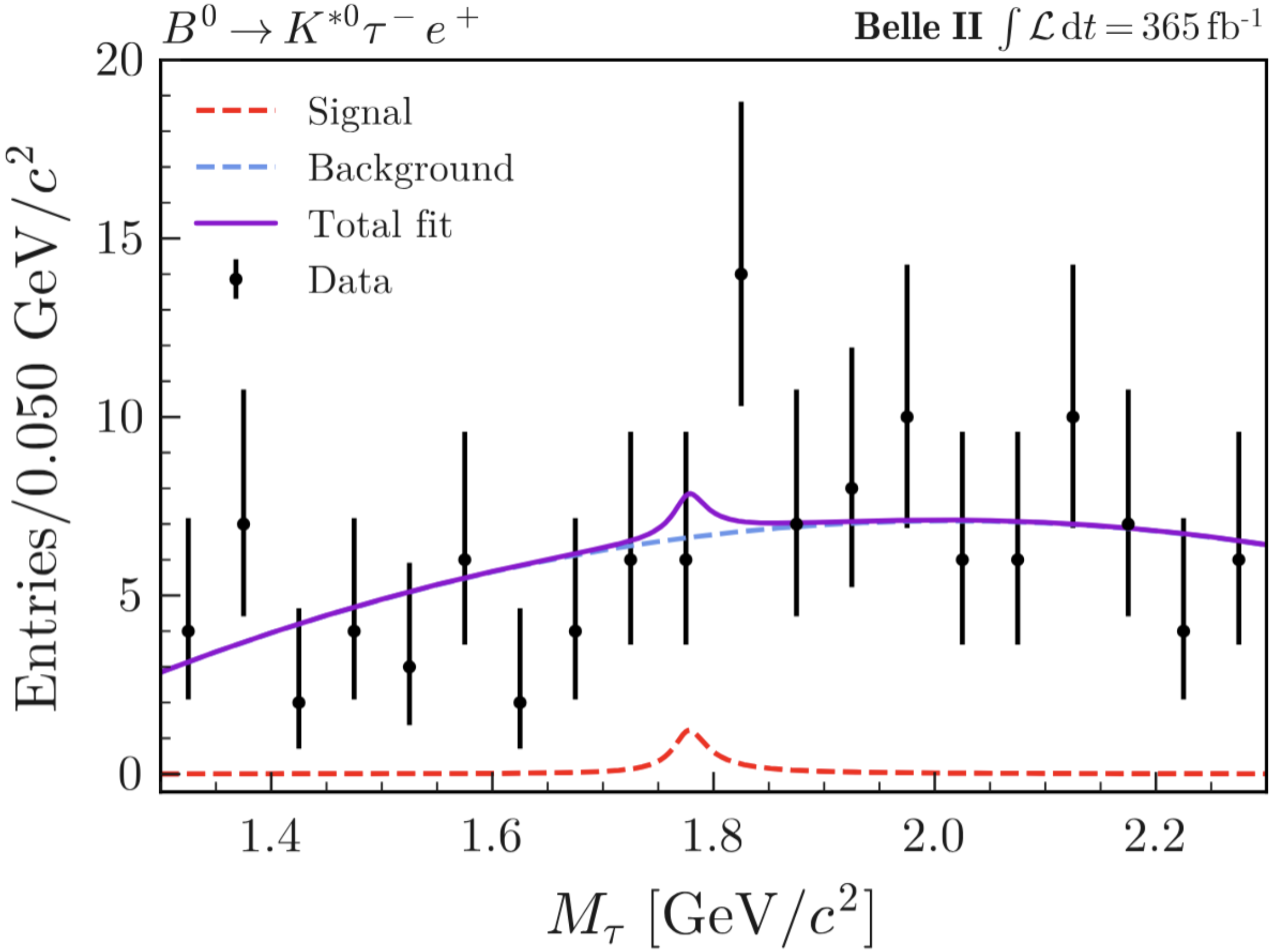}\label{fig:b3}}
\caption{Fits to the recoil mass in the $\bar{b} \to \bar{s}\tau^- e^+$ data for the
\protect\subref{fig:a2}: $K_S^0$ mode, where Belle and Belle~II data are combined (from~\cite{KStaul});
\protect\subref{fig:b2},\protect\subref{fig:b3}: $K^{*0}$ mode, with a simultaneous fit to the two datasets (from~\cite{K*0taul}).
\label{fig:stauell}}
\end{figure}

These proceedings focus on recent searches for $B^0 \to K_S^0\tau \ell$~\cite{KStaul} and $B^0 \to K^{*0}\tau \ell$~\cite{K*0taul} using combined Belle and Belle~II datasets with hadronic $B$-tagging. The signal appears as a peak at the $\tau$ mass in the recoil mass, which offers excellent resolution and smooth backgrounds without peaking components. Sidebands of the recoil mass allow to validate the simulation used to train background-suppressing classifiers. We exploit event-shape variables for $q\bar{q} \;(q=u,d,s,c)$ background suppression, and kinematic variables to reduce the $B\bar{B}$ component; for example, the invariant mass of the kaon and $\ell$ (or the $\tau$ daughter, depending on the considered $b \to s\tau^\pm\ell^\mp$ charge configuration), effective against $D$-meson background. $B^0 \to D^-D_s^+$ decays with similar topology compared to $B^0 \to K^{(*)0}\tau\ell$ are used to calibrate the signal shape and the BDT cut efficiency. Fits to data for the $K_S^0$ and $K^{*0}$ channels in a representative $\tau\ell$ configuration are shown in Fig.~\ref{fig:stauell}. Selection efficiency maps as functions of the relevant degrees of freedom are provided in Refs.~\cite{KStaul,K*0taul} to enable reinterpretation, with the two modes being sensitive to different NP mediators.

\section{Conclusion}
Electroweak penguin and lepton-flavor-violating $B$~decays provide sensitive probes of the Standard Model and beyond. Recent results from Belle and Belle~II demonstrate the competitiveness, and in some cases the uniqueness, of $B$-factories compared to other flavor experiments~\cite{esppu}. While analyzing and accumulating more Run~2 data, Belle~II is refining its methodologies and exploring new approaches to enhance sensitivity in $B$~decays with missing energy, as well as establishing frameworks for the quantitative reinterpretation of its results.

\section*{Acknowledgments}
This project has received funding from the European Union's Horizon 2020 research and innovation program under the ERC grant agreement No. \href{https://cordis.europa.eu/project/id/884719}{884719}.

\bibliographystyle{style}
\bibliography{references}

\end{document}